\documentclass[pra, twocolumn, showpacs, preprintnumbers, amsmath, amssymb,
superscriptaddress, aps]{revtex4}%
\usepackage{hyperref}
\usepackage{amssymb}
\usepackage{latexsym}
\usepackage{epsfig}
\usepackage{dcolumn}
\usepackage{amsmath}
\usepackage{amsfonts}
\usepackage{bm}
\usepackage{color}
\usepackage{graphicx}%
\providecommand{\U}[1]{\protect\rule{.1in}{.1in}}

\begin{document}
%
\begin{abstract}
In this paper we calculate the nonlinear susceptibility and the resonant Raman
cross section for the paramagnetic phase of the ferromagnetic Quantum Ising
model in one dimension. In this region the spectrum of the Ising model has a
gap $m$. The Raman cross section has a strong singularity when the energy of
the outgoing photon is at the spectral gap $\omega_{f} \approx m$ and a square
root threshold when the frequency difference between the incident and outgoing
photons $\omega_{i} -\omega_{f} \approx 2m$. The latter feature reflects the
fermionic nature of the Ising model excitations.
\end{abstract}
\title
{Probing Strong Correlations with Light Scattering: the Example of the Quantum
Ising model}
\author{H.~M.~Babujian}
\affiliation
{Yerevan Physics Institute, Alikhanian Brothers 2, Yerevan, 375036 Armenia and
International Institute of Physics, Universidade Federal do Rio Grande do
Norte (UFRN), 59078-400 Natal-RN, Brazil, Simons Center for Geometry and
Physics, Stony Brook University, USA}
\author{M.~Karowski}
\affiliation{Institut f\"{u}r Theoretische Physik, Freie Universit\"{a}%
t Berlin, Arnimallee
14, 14195 Berlin, Germany}
\author{A.~M.~Tsvelik}
\affiliation
{Condensed Matter Physics and Materials Science Division, Brookhaven National
Laboratory, Upton, NY 11973-5000, USA}
\maketitle

{\bf Introduction.} When photons enter into a strongly interacting medium one expects nonlinear
phenomena such as frequency mixing and inelastic light scattering. These
effects are interesting by itself, but they can also serve as experimental
tools to extract otherwise unaccesible information about strongly correlated
dynamics. All these tools probe multi-point dynamical correlation functions
which carry much richer information than a simple linear response. In this
paper we will discuss three- and four-point dynamical correlation functions
for the ferromagnetic Quantum Ising (FQI) model in one spatial dimension (1D)
and relate them to two spectroscopic probes: the nonlinear susceptibility and
the inelastic scattering cross section of light ( Raman scattering). We have
chosen the 1D FQI model for three reasons. Firstly, this is a strongly
correlated model whose applicability to real materials has been firmly
established (see the discussion at the end of the paper). Secondly, this model
admits a resonance regime of where nonlinear effects are strongly enhanced.
The third factor is a comparative simplicity of the calculations which allows
us not to delve into much technical details.


The FQI model is described by the Hamiltonian
\begin{equation}
H=\sum_{n}\Big(-J\sigma_{n}^{z}\sigma_{n+1}^{z}+h\sigma^{x}\Big),
\label{model}%
\end{equation}
where $\sigma^{a}$ are the Pauli matrices. This model has numerous condensed
matter realizations being one of the most popular models of condensed matter
theory. It describes a sequence of coupled two level systems. They may
represent spins; then the first term describes an anisotropic exchange
interaction. In this case $\sigma^{z}$ directly couples to external magnetic
field: $\mu_{B}B_{n}^{z}\sigma_{n}^{z}$.

States of the two level systems may also correspond to positions of electric
charges in a double well potential. Then the first term is the dipole-dipole
interaction and the transverse field describes the quantum tunneling between
the wells. Then $\sigma^{a}$ would be the dipole moment operators. Their
interaction with the electric field is given by $pE_{n}^{z}\sigma_{n}^{z}$
with $p$ being the dipole moment.

Since the dominant interaction is ferromagnetic, the strongest fluctuations
take place at zero wave vectors which guarantees a direct coupling to the
electromagnetic field creating optimal resonance conditions. The Ising model
(\ref{model}) has two phases depending on the sign of $m=h-J$. The resonance
occurs in the paramagnetic phase $m>0$ when the ground state average of the
order parameter $\langle\,\sigma^{z}\,\rangle=0$. In that case the
electromagnetic field has a nonzero matrix element between the ground state
and single magnon state.

Raman light scattering is a powerful experimental technique frequently used in
condensed matter physics. The measured quantity is the inelastic scattering
cross section of photons $R(\mathbf{q},\Omega)$ which contains information
about the excitations of condensed matter systems with which the photons
interact. The theory of Raman scattering was formulated in the nineteen
twenties \cite{kramers,dirac} when the formulae for $R(\mathbf{q},\Omega)$
were derived (see also \cite{shraiman}). A radical simplification of these
formulae was suggested in \cite{shvaika} where the resonant part of the Raman
cross section was expressed as a particular limit of the four point
correlation function of the current operators. This simplification allows one
to apply to the problem various techniques of quantum field theory such as
Feynman diagram expansion and also simplifies the application of
nonperturbative techniques.

Another technique to be discussed is a nonlinear response directly related to
the three-point correlation function. The related themes are two-dimensional
spectroscopy and spectroscopy with entangled photons \cite{cundiff}%
,\cite{mukamel}

The Jordan-Wigner transformation transforms FQI into a model of noninteracting
noninteracting massive Majorana fermions. In the scaling limit $m<<J$ their
dispersion becomes relativistic $\epsilon(k)=\sqrt{v^{2}k^{2}+m^{2}},~~v\sim
J.$ In what follows we will set $v=1$ to restore $J$ in the final expressions.

The fact that the excitations of the Ising model do not interact does not
make the model trivial. Indeed, since $\sigma^{z}$ operators are very nonlocal
in terms of the fermions, the electromagnetic field has matrix elements
between states with different number of fermionic excitations. Such situation
is typical for strongly interacting systems and experimental probes of
multipoint correlators are highly suitable to reveal this nonlocality. In the
paramagnetic phase of FQI the inelastic processes involve matrix elements with
odd number of the Ising fermions with the leading low energy processes being
transitions from single- to two-fermion states. The fermionic nature of the
excitations is reflected in the fact that the cross section vanishes at the
threshold: $R(\Omega)\sim(\Omega-2m)^{1/2}$ (see Eqs. (\ref{r1},\ref{r2}) below.


\subparagraph{The nonlinear susceptibilit\textbf{y} }

is the third derivative of the action with respect to the dynamical magnetic
field. In the Ising model such response exists only when the polarization of
photons is such that the magnetic field has both $z$ and $x$ components:
\begin{align}
&  V=\mu_{B}\sum_{n}(B^{z}\sigma_{n}^{z}+B^{x}\sigma_{n}^{x}),\nonumber\\
&  \chi_{zzx}^{(3)}=\langle\hat{T}\sigma^{z}(t_{1})\sigma^{z}(t_{2})\sigma
^{x}(t_{3})\rangle_{connected},\label{three}%
\end{align}
where $\sigma=\sum_{n}\sigma_{n}^{z}$ (we assume that electromagnetic
radiation carries no momentum). The nonlinear susceptibility describes the
effects or frequency mixing.

\subparagraph{Relation between the Raman cross section and the correlation
functions:}

As it was stated above we assume the following spin-photon interaction
$V=\sum_{n}pE_{n}\sigma_{n}^{z}$. Then according to Eqs. (2.21) from
\cite{shvaika} the cross section for the light beam polarized along the
$z$-axis is given by the following expression
\begin{equation}
R(\omega_{i},\omega_{f})=2\pi{\mu_{B}}^{4}[(h\omega_{i})(h\omega_{f}%
)]\frac{\chi_{R}(\omega_{i},\omega_{f})}{1-\exp[-\beta(\omega_{i}-\omega
_{f})]},\label{raman}%
\end{equation}
where $\beta=1/T$, $\omega_{i}$ and $\omega_{f}$ are frequencies of the
incident and the scattered light. We will consider the $T=0$ limit. Then the
function $\chi$ is expressed as (see Eqs. 2.30, 2.31 from \cite{shvaika})
\begin{multline}
\chi_{R}(\omega_{i},\omega_{f})=\frac{1}{2\pi i}\,\lim_{\delta_{1}>\delta
_{2}\rightarrow0}\label{chi}\\
\Big\{\tilde{\Xi}(-\omega_{i}-i\delta_{1},\omega_{f}+i\delta_{2},-\omega
_{f}+i\delta_{2},\omega_{i}-i\delta_{1})-\\
\tilde{\Xi}(-\omega_{i}-i\delta_{2},\omega_{f}+i\delta_{1},-\omega_{f}%
+i\delta_{1},\omega_{i}-i\delta_{2})\Big\},
\end{multline}
where $\tilde{\Xi}$ is the Fourier transform of the four-point time ordered
correlation function:
\begin{equation}
\langle\,\hat{T}\sigma(t_{1})\sigma(t_{2})\sigma(t_{3})\sigma(t_{4}%
)\,\rangle_{connected}\label{Gamma}%
\end{equation}
up to the energy momentum $\delta$-function. Below we will derive the
expression for (\ref{Gamma}) in the paramagnetic phase of model (\ref{model})
at $T=0$ in the limit $m=h-J<<J$ and will use the result to calculate the
Raman cross section (\ref{raman}).



\subparagraph{The results:}

The only nonzero third order response (\ref{three}) includes two magnetic
fields with frequencies $\omega_{z},\omega_{x}-\omega_{z}$ along the $z$- and
one field with frequency $-\omega_{x}$ along the $x$-direction. Our result
where we take into account only two magnon production processes, is
\begin{multline}
\chi_{zzx}^{(3)}(\omega_{z},-\omega_{z}+\omega_{x},-\omega_{x})= C^2 \mu_B^3(mJ^3)^{-1/4}\times\\
\Big\{\frac{\left(  \omega_{x}-2m\right)   h\left(  \omega_{x}
/2m\ + i\delta\right)   }{\left(  \omega_{z}-m\right)  \left(
\omega_{x}-\omega_{z}-m\right)  }+\left(  \omega_{x,z}\rightarrow-\omega
_{x,z}\right)  \\
+16m\pi\frac{\left(  \omega_{x}-\omega_{z}\right)  \omega_{z}-m^{2}}{\left(
\omega_{z}^{2}-m^{2}\right)  (\left(  \omega_{x}-\omega_{z}\right)  ^{2}%
-m^{2})}\Big\}\label{threeP}%
\end{multline}
where $C\sim 1$ is a numerical constant and (see \cite{SM})
\[
h\left(  x\right)  =\int_{-\infty}^{\infty}\left(  1+\frac{1}{\cosh\theta
}\right)  ^{2}\frac{1}{\cosh\theta-x}d\theta.
\]
For the Raman function we obtain for $\omega_{i}>\omega_{f}>0$ substituting
eq. (\ref{G1}) into (\ref{chi}) (see \cite{SM})
\begin{multline}
\chi_{R}(\omega_{i},\omega_{f})\sim(m/J)^{1/2}\frac{J^{3}}{m^{4}}%
\Big[\frac{G\left(  \frac{1}{2m}(\omega_{i}+\omega_{f})\right)  }{\left(
\omega_{f}-m\right)  ^{2}\left(  \omega_{i}-m\right)  ^{2}}\label{r1}\\
+\frac{G\left(  \frac{1}{2m}(\omega_{i}-\omega_{f})\right)  }{\left(
\omega_{f}+m\right)  ^{2}\left(  \omega_{i}-m\right)  ^{2}}\Big]
\end{multline}
where
\begin{equation}
G(x)=\Theta(x-1)\frac{\left(  x-1\right)  ^{1/2}\left(  x+1\right)  ^{5/2}%
}{x^{3}}\,.\label{r2}%
\end{equation}
The calculation takes into account only 2-particle intermediate states which
is allowed in the range of frequencies $\left\vert \omega_{i}-\omega
_{f}\right\vert <4m$ when the processes with emission of more than 2 particles
do not contribute to the inelastic cross section.
The threshold for the inelastic scattering is at $\omega_{i}\pm\omega_{f}=2m$
corresponding to the emission of two fermionic excitations.

\subparagraph{Green's functions:}

Below we will do our calculations in the most general form valid for all
integrable models and at the end apply the results to the Ising model. We will
concentrate on the most difficult case of the four-point function, the
calculations of the three-point one are comparatively straightforward.

The Green's functions are time ordered n-point functions%
\begin{align*}
\tau(\underline{x}) &  =\langle\,0\,|\,T\varphi(x_{1})\,\dots\,\varphi
(x_{n})|\,0\,\rangle\\
&  =\sum_{perm(x)}\Theta_{1\dots n}(\underline{t})w(\underline{x})\,,
\end{align*}
here $w(\underline{x})=\langle\,0\,|\,\varphi(x_{1})\,\dots\,\varphi
(x_{n})|\,0\,\rangle$ is the Wightman function and $\Theta_{1\dots
n}(\underline{t})=\Theta(t_{12})\Theta(t_{23})\dots\,\Theta(t_{n-1,n})$. In
momentum space%
\begin{align*}
\tilde{\tau}(\underline{k}) &  =\int\underline{d^{2}x}e^{ix_{i}k_{i}}%
\tau(\underline{x})\\
&  =\sum_{perm(k)}\int\underline{d^{2}x}e^{ix_{i}k_{i}}\Theta_{1\dots
n}\,w(\underline{x})\,.
\end{align*}
The connected Green's functions are given by%
\begin{equation}
\tilde{\tau}(\underline{k})=\sum_{\underline{k}_{1}\cup\dots\cup\underline
{k}_{m}=\underline{k}}\tilde{\tau}_{c}(\underline{k}_{1})\dots\tilde{\tau}%
_{c}(\underline{k}_{m})\,.\label{tauc0}%
\end{equation}
For convenience we split off the energy momentum $\delta$-function and define
$\tilde{\Pi}(\underline{k})$by%
\begin{equation}
\tilde{\tau}_{c}(\underline{k})=\left(  2\pi\right)  ^{2}\delta^{(2)}\left(
{\textstyle\sum}
k_{i}\right)  \,i\,\tilde{\Pi}(\underline{k})\,.\label{G}%
\end{equation}

\textbf{S-matrix and form factors. } For integrable quantum field theories the
n-particle S-matrix factorizes into $n(n-1)/2$ two-particle ones
\[
S^{(n)}(\theta_{1},\dots,\theta_{n})=\prod_{i<j}S(\theta_{ij})\,,
\]
where the product on the right hand side has to be taken in a specific order
(see e.g.~\cite{KT}). The numbers $\theta_{ij}$ are the rapidity differences
$\theta_{ij}=\theta_{i}-\theta_{j}$, which are related to the momenta of the
particles by $p_{i}=m\left(  \cosh\theta_{i},\sinh\theta_{i}\right)  $. The
form factors of a bosonic field are the matrix elements%
\begin{equation}
F(\underline{\theta})=\langle\,0\,|\,\varphi(0)\,|\,\theta_{1},\dots
,\theta_{n}\,\rangle\, \label{F0}%
\end{equation}
For the paramagnetic phase they are non-zero for $n=$ odd. They satisfy the
form factor equations (i) -- (v) (see e.g. \cite{BFK}). We use the
normalization $\langle\,0\,|\,\varphi(0)\,|\,\theta\,\rangle=1$. As a
generalization we write
\begin{equation}
F(\underline{\theta}^{\prime};\underline{\theta})=\langle\,\theta_{n^{\prime}%
}^{\prime},\dots,\theta_{1}^{\prime}\,|\,\varphi(0)\,|\,\theta_{1}%
,\dots,\theta_{n}\,\rangle\label{F}%
\end{equation}
which is related to (\ref{F0}) by crossing. In particular (see \cite{SM})%
\begin{align}
&  F(\theta_{1};\theta_{2},\theta_{3})=F(\theta_{1},\theta_{2}-i\pi_{-}%
,\theta_{3}-i\pi_{+})+\delta_{\theta_{12}}+\delta_{\theta_{13}}\label{F12}\\
&  F(\theta_{2},\theta_{3};\theta_{4})=F(\theta_{3}+i\pi_{+},\theta_{2}%
+i\pi_{-},\theta_{4})+\delta_{\theta_{24}}+\delta_{\theta_{34}} \label{F21}%
\end{align}
with $i\pi_{\pm}=i\pi\pm i\epsilon$ and $\delta_{\theta_{12}}=4\pi
\delta(\theta_{1}-\theta_{2})$.

\subparagraph{The Green's functions in low particle approximation:}

The 2-point Wightman function in 1-particle intermediate states approximation
is (with the short notation $\int_{\theta}=\frac{1}{4\pi}\int d\theta$)%
\begin{multline*}
w^{1}(x_{1}-x_{2})=\int_{\theta}\langle\,0\,|\,\varphi(x_{1})\,|\,\theta
\,\rangle\langle\,\theta\,|\,\varphi(x_{2})\,|\,0\rangle\\
=\int\frac{dp}{2\pi2\omega}e^{-i(x_{1}-x_{2})p}=i\Delta_{+}\left(  x_{1}%
-x_{2}\right)  \,.
\end{multline*}
The 4-point Wightman function in 1-0-1 intermediate particle approximation is
\begin{multline}
w^{101}(\underline{x})=\int_{\theta_{1}}\langle\,0\,|\,\varphi(x_{1}%
)\,|\,\theta_{1}\,\rangle\langle\,\theta_{1}\,|\,\varphi(x_{2})\,|\,0\rangle\\
\times\,\int_{\theta_{4}}\langle\,0\,|\,\varphi(x_{3})\,|\,\theta_{4}%
\rangle\langle\,\theta_{4}\,|\,\varphi(x_{4})\,|\,0\,\rangle\\
=w^{1}(x_{1}-x_{2})w^{1}(x_{3}-x_{4})\,.\label{w0}%
\end{multline}
The 4-point Wightman function in 1-2-1 intermediate particle approximation is
(with $\int_{\underline{\theta}}=\int_{\theta_{1}}\dots\int_{\theta_{4}}$ and
$\mathbf{xp}=x_{1}p_{1}+x_{2}\left(  p_{2}+p_{3}-p_{1}\right)  +x_{3}\left(
p_{4}-p_{2}-p_{3}\right)  -x_{4}p_{4}$)
\begin{multline*}
w^{121}(\underline{x})=\frac{1}{2}\int_{\underline{\theta}}\langle
\,0\,|\,\varphi(x_{1})\,|\,\theta_{1}\,\rangle\,\langle\,\theta_{1}%
\,|\,\varphi(x_{2})\,|\,\theta_{2},\theta_{3}\rangle\\
\times\,\langle\,\theta_{3},\theta_{2}\,|\,\varphi(x_{3})\,|\,\theta
_{4}\rangle\,\langle\,\theta_{4}\,|\,\varphi(x_{4})\,|\,0\,\rangle\\
=\frac{1}{2}\int_{\underline{\theta}}e^{-i\mathbf{xp}}F(\theta_{1};\theta
_{2},\theta_{3})F(\theta_{2},\theta_{3};\theta_{4})
\end{multline*}
with (see (\ref{F12}) and (\ref{F21}))
\begin{align*}
&  \tfrac{1}{2}F(\theta_{1};\theta_{2},\theta_{3})F(\theta_{2},\theta
_{3};\theta_{4})\\
&  =\tfrac{1}{2}\left(  F(\theta_{1},\theta_{2}-i\pi_{-},\theta_{3}-i\pi
_{+})+\delta_{\theta_{12}}+\delta_{\theta_{13}}\right)  \\
&  \times\left(  F(\theta_{3}+i\pi_{+},\theta_{2}+i\pi_{-},\theta_{4}%
)+\delta_{\theta_{24}}+\delta_{\theta_{34}}\right)  \\
&  =I(\underline{\theta})=I_{1}(\underline{\theta})+I_{2}(\underline{\theta
})+I_{3}(\underline{\theta})\,.
\end{align*}
We have introduced (see \cite{SM})%
\begin{align}
I_{1}(\underline{\theta}) &  =\tfrac{1}{4}F(\theta_{1},\theta_{2}-i\pi
_{+},\theta_{3}-i\pi_{-})\label{I}\\
&  \times F(\theta_{3}+i\pi_{+},\theta_{2}+i\pi_{-},\theta_{4})\nonumber\\
&  +\tfrac{1}{4}F(\theta_{1},\theta_{2}-i\pi_{-},\theta_{3}-i\pi
_{+})\nonumber\\
&  \times F(\theta_{3}+i\pi_{-},\theta_{2}+i\pi_{+},\theta_{4})\nonumber\\
I_{2}(\underline{\theta}) &  =\tfrac{1}{4}\left(  \delta_{\theta_{12}}\left(
1+S(\theta_{23})\right)  +\delta_{\theta_{13}}\left(  1+S(\theta_{23})\right)
\right)  \nonumber\\
&  \times F(\theta_{3}+i\pi_{+},\theta_{2}+i\pi_{-},\theta_{4})\nonumber\\
&  +\tfrac{1}{4}F(\theta_{1},\theta_{2}-i\pi_{-},\theta_{3}-i\pi
_{+})\nonumber\\
&  \times\left(  \delta_{\theta_{24}}\left(  1+S(\theta_{32})\right)
+\delta_{\theta_{34}}\left(  1+S(\theta_{32})\right)  \right)  \nonumber\\
I_{3}(\underline{\theta}) &  =\tfrac{1}{2}\left(  \delta_{\theta_{12}}%
+\delta_{\theta_{13}}\right)  \left(  \delta_{\theta_{24}}+\delta_{\theta
_{34}}\right)  \,.\nonumber
\end{align}
 From $I_{3}$ we calculate%
\begin{multline}
w_{3}^{121}(\underline{x})=w^{1}\left(  x_{1}-x_{4}\right)  w^{1}\left(
x_{2}-x_{3}\right)  \label{w3}\\
+w^{1}\left(  x_{1}-x_{3}\right)  w^{1}\left(  x_{2}-x_{4}\right)  \,.
\end{multline}
Therefore neglecting contributions from higher particle intermediate states
using (\ref{tauc0}) and (\ref{w0}) we obtain the connected 4-point Green's
function%
\begin{multline}
\tilde{\tau}_{c}(\underline{k})=\sum_{perm(k)}\int\underline{d^{2}x}%
\Theta_{1\dots n}e^{ix_{i}k_{i}}\label{tauc1}\\
\times\left(  w_{1}^{121}(\underline{x})+w_{2}^{121}(\underline{x})\right)
\end{multline}
where $w_{i}^{121}(\underline{x})$ is given by the contribution from
$I_{i}(\underline{\theta})$ in (\ref{I}). For $k_{i}=(k_{i}^{0},0)$ we obtain
from (\ref{G}) (see \cite{SM})
\begin{align}
\tilde{\Xi}(\underline{k}) &  =\frac{1}{32\pi m^{6}}\sum_{perm(k)}\label{G1}\\
&  \times\frac{m}{m-k_{1}^{0}-i\epsilon}\frac{m}{k_{4}^{0}+m-i\epsilon
}g\left(  \frac{-1}{2m}\left(  k_{3}^{0}+k_{4}^{0}\right)  \right)  \nonumber
\end{align}
with%
\begin{equation}
g(x)=-2\pi\int_{\theta}\frac{1}{\omega/m}I(0,\theta,-\theta,0)\frac{1}%
{\omega/m-x}\label{g}%
\end{equation}
where $I=I_{1}+I_{2}$ contribute. For integrable models typically $S(0)=-1$,
then the contribution from $I_{2}$ vanishes for $\theta_{i}\rightarrow0$. With
(\ref{chi}) and $G(x)=(x-1)^{2}\operatorname{Im}g(x)$ equation (\ref{r1})
follows. Next we consider a simple model, for which we calculate the function
$g(x)$ explicitly.

\subparagraph{The scaling Ising model:}

In the scaling limit this model may be described by an interacting Bose field
$\sigma_{n}^{z}=Cm^{1/8}\phi(x)$, where $C$ is a numerical constant and
$m=h-J$. The excitations are noninteracting Majorana fermions with the
2-particle S-matrix $S(\theta)=-1$. The field $\sigma^{x}=(m/J)^{1/2}%
\epsilon(x)\sim\bar{\psi}\psi(x)$, where $\psi$ is a free Majorana spinor
field. In \cite{BKW,K2} the form factor was proposed%
\begin{equation}
F(\underline{\theta})=\langle\,0|\,\sigma(0)\,|\,\theta_{1},\dots,\theta
_{n}\,\rangle=\left(  2i\right)  ^{\frac{n-1}{2}}\prod_{i<j}\tanh\tfrac{1}%
{2}\theta_{ij}.\label{Fn}%
\end{equation}
For $k_{i}=(k_{i}^{0},0)$ in momentum space the contribution from $I_{2}$ in
(\ref{I}) vanishes, because $S(0)=-1$. From (\ref{I}) and (\ref{Fn}) we obtain
(see \cite{SM})
\[
I_{1}(0,\theta,-\theta,0)=\tanh^{2}\theta\coth^{4}\tfrac{1}{2}\left(
\theta-i\epsilon\right)  +\left(  \epsilon\rightarrow-\epsilon\right)  .
\]
Substituting it into (\ref{g}) and taking into account $G(x)=(x-1)^{2}%
\operatorname{Im}g(x)$ and the relation between $\sigma^{z}$ and $\phi$ we
obtain (\ref{r2}).

>From $\epsilon(x)\sim\bar{\psi}\psi(x)$ one has for a free Majorana spinor
field
\begin{equation}
\langle0|\epsilon(0)|\theta_{1},\theta_{2}\rangle=\sinh(\theta_{12}%
/2).\label{e2}%
\end{equation}
For low intermediate particle numbers this leads to (\ref{threeP}) as above
(see \cite{SM}).


\subparagraph{Conclusions:}

 We calculated the three and the four point correlation functions for the
ferromagnetic Quantum Ising model and discussed their relation to the
observable quantities. In the paramagnetic phase of FQI the magnetic field is
directly coupled to the spin operator which has matrix elements between states
with odd and even number of the Ising fermions. The fact that light can create
odd number of fermionic excitations is quite remarkable. It emphasizes an
ambiguity between bosons and fermions existing in one dimension. 

The best experimental realization of FQI model known to date is found in
columbite CoNb$_{2}$O$_{6}$ \cite{coldea},\cite{ong},\cite{armitage}. Another
possible candidate is Sr$_{3}$CuIrO$_{6}$ \cite{yin}. Both these materials are
quasi 1D insulators; the columbite displays a quantum critical point at
$B=5.5$T which is very well described by the theory of the Ising model
\cite{ong}. Neutron scattering \cite{coldea} and terahertz spectroscopy
\cite{armitage} also yield excellent agreement with the theoretical
predictions. In the view of these we suggest that a good test of our theory
would be high field spectroscopic measurements at terahertz frequencies on
CoNb$_{2}$O$_{6}$.

\subparagraph{\textbf{Acknowledgments:}}

We are grateful to G. Blumberg, J. Misewich and especially to N. P. Armitage
for advising us on the experimentally related matters, to S. Lukyanov who
pointed out for us paper \cite{BNNPSW} and to A. B. Zamolodchikov for fruitful
discussions. A.~M.~T. was supported by the U.S. Department of Energy (DOE),
Division of Materials Science, under Contract No. DE-AC02-98CH10886. H.~B. is
grateful to Simons Center and Brookhaven National Laboratory for hospitality
and support. H.~B. also supported by Armenian grant 15T-1C308 and by ICTP
OEA-AC-100 project. M.~K. was supported by Fachbereich Physik, Freie
Universit\"{a}t Berlin.


\widetext

\bigskip

\begin{center}
\textbf{{\large Supplementary Material: }}
\end{center}

\setcounter{equation}{0} \setcounter{figure}{0} \setcounter{table}{0}
\setcounter{page}{1}
\makeatletter\renewcommand{\theequation}{S\arabic{equation}}
\renewcommand{\thefigure}{S\arabic{figure}}
\renewcommand{\bibnumfmt}[1]{[S#1]} \renewcommand{\citenumfont}[1]{S#1}
\makeatother

\underline{\textbf{Proof of (\ref{r1})}}

\subparagraph{Crossing:}

The form factors (\ref{F}) satisfy crossing relations (see e.g. (31) in
\cite{BK} ), in particular
\begin{align}
F(\theta_{1};\theta_{2},\theta_{3}) &  =\,\langle\,\theta_{1}\,|\,\varphi
(0)\,|\,\theta_{2},\theta_{3}\rangle=F(\theta_{1}+i\pi_{-},\theta_{2}%
,\theta_{3})+\delta_{\theta_{12}}+\delta_{\theta_{13}}S(\theta_{23}%
)\label{cr}\\
F(\theta_{2},\theta_{3};\theta_{4}) &  =\,\langle\,\theta_{3},\theta
_{2}\,|\,\varphi(0)\,|\,\theta_{4}\rangle=F(\theta_{3}+i\pi_{-},\theta
_{2}+i\pi_{-},\theta_{4})+\delta_{\theta_{24}}+\delta_{\theta_{34}}%
S(\theta_{32})\nonumber
\end{align}
with $i\pi_{\pm}=i\pi\pm i\epsilon$ and $\delta_{\theta_{12}}=4\pi
\delta(\theta_{1}-\theta_{2})$. Using the form factor equation (iii) and
Lorentz invariance (see e.g. \cite{BFK})
\begin{align*}
\operatorname*{Res}_{\theta_{12}=i\pi}F(\theta_{1},\theta_{2},\theta_{3}) &
=2i\,\left(  \mathbf{1}-S(\theta_{23})\right)  \\
F(\theta_{1},\theta_{2},\theta_{3}) &  =F(\theta_{1}+\mu,\theta_{2}+\mu
,\theta_{3}+\mu)
\end{align*}
we can rewrite these equations as (\ref{F12}) and (\ref{F21}). And further one
derives%
\begin{align}
\tfrac{1}{2}F(\theta_{1},\theta_{2}-i\pi_{-},\theta_{3}-i\pi_{+}%
)+\delta_{\theta_{12}}+\delta_{\theta_{13}} &  =\tfrac{1}{2}\left(
F(\theta_{1},\theta_{2}-i\pi_{+},\theta_{3}-i\pi_{-})+\delta_{\theta_{12}%
}\left(  1+S(\theta_{23})\right)  +\delta_{\theta_{13}}\left(  1+S(\theta
_{23})\right)  \right)  \label{F12a}\\
\tfrac{1}{2}F(\theta_{3}+i\pi_{+},\theta_{2}+i\pi_{-},\theta_{4}%
)+\delta_{\theta_{24}}+\delta_{\theta_{34}} &  =\tfrac{1}{2}\left(
F(\theta_{3}+i\pi_{-},\theta_{2}+i\pi_{+},\theta_{4})+\delta_{\theta_{24}%
}\left(  1+S(\theta_{32})\right)  +\delta_{\theta_{34}}\left(  1+S(\theta
_{32})\right)  \right)  \,.\label{F21a}%
\end{align}
Using equations (\ref{F12},\ref{F21}) and the identity

$\left(  a+b+c\right)  \left(  d+e+f\right)  =\left(  \tfrac{1}{2}%
a+b+c\right)  d+a\left(  \tfrac{1}{2}d+e+f\right)  +\left(  b+c\right)
\left(  e+f\right)  $ we derive%
\begin{align*}
&  F(\theta_{1};\theta_{2},\theta_{3})F(\theta_{2},\theta_{3};\theta_{4})\\
&  =\left(  F(\theta_{1},\theta_{2}-i\pi_{-},\theta_{3}-i\pi_{+}%
)+\delta_{\theta_{12}}+\delta_{\theta_{13}}\right)  \left(  F(\theta_{3}%
+i\pi_{+},\theta_{2}+i\pi_{-},\theta_{4})+\delta_{\theta_{24}}+\delta
_{\theta_{34}}\right) \\
&  =\left(  \tfrac{1}{2}F(\theta_{1},\theta_{2}-i\pi_{-},\theta_{3}-i\pi
_{+})+\delta_{\theta_{12}}+\delta_{\theta_{13}}\right)  F(\theta_{3}+i\pi
_{+},\theta_{2}+i\pi_{-},\theta_{4})\\
&  +F(\theta_{1},\theta_{2}-i\pi_{-},\theta_{3}-i\pi_{+})\left(  \tfrac{1}%
{2}F(\theta_{3}+i\pi_{+},\theta_{2}+i\pi_{-},\theta_{4})+\delta_{\theta_{24}%
}+\delta_{\theta_{34}}\right) \\
&  +\left(  \delta_{\theta_{1}\theta_{2}}+\delta_{\theta_{1}\theta_{3}%
}\right)  \left(  \delta_{\theta_{4}\theta_{2}}+\delta_{\theta_{4}\theta_{3}%
}\right)
\end{align*}
then (\ref{F12a}) and (\ref{F21a}) prove (\ref{I}).

\subparagraph{Calculation of $\tau_{c}^{121}$:}

To derive (\ref{G1}) from (\ref{tauc1}) we calculate (for $i=1,2$)%
\begin{align*}
&  \int\underline{d^{2}x}\Theta_{1\dots n}e^{ix_{i}k_{i}}\int_{\underline{p}%
}e^{-ix_{1}p_{1}-ix_{2}\left(  p_{2}+p_{3}-p_{1}\right)  -ix_{3}\left(
p_{4}-p_{2}-p_{3}\right)  +ix_{4}p_{4}}I_{i}(\theta_{1},\theta_{2},\theta
_{3},\theta_{4})\\
&  =\int\underline{dx}^{0}\Theta_{1\dots n}e^{ix_{i}^{0}k_{i}^{0}}%
\int_{\underline{\theta}}e^{-ix_{1}^{0}\omega_{1}-ix_{2}^{0}\left(  \omega
_{2}+\omega_{3}-\omega_{1}\right)  -ix_{3}^{0}\left(  \omega_{4}-\omega
_{2}-\omega_{3}\right)  +ix_{4}^{0}\omega_{4}}\\
&  \left(  2\pi\right)  ^{4}\delta(p_{1}-k_{1}^{1})\delta(p_{2}+p_{3}%
-p_{1}-k_{2}^{1})\delta(p_{4}-p_{2}-p_{3}-k_{3}^{1})\delta(p_{4}+k_{4}%
^{1})I_{i}(\theta_{1},\theta_{2},\theta_{3},\theta_{4}).
\end{align*}
For $k_{i}=(k_{i}^{0},0)$ this is equal to%
\begin{align*}
&  =2\pi\delta\left(  k_{1}^{1}+k_{2}^{1}+k_{3}^{1}+k_{4}^{1}\right)
\int\underline{dx}^{0}\Theta_{1\dots n}e^{ix_{i}^{0}k_{i}^{0}}\int
_{\underline{\theta}}e^{-ix_{1}^{0}m-ix_{2}^{0}\left(  \omega_{2}+\omega
_{3}-m\right)  -ix_{3}^{0}\left(  m-\omega_{2}-\omega_{3}\right)  +ix_{4}%
^{0}m}\\
&  \times\left(  2\pi\right)  ^{3}\delta(p_{1})\delta(p_{2}+p_{3})\delta
(p_{4})I_{i}(0,\theta_{2},-\theta_{2},0)\\
&  =2\pi\delta\left(  k_{1}^{1}+k_{2}^{1}+k_{3}^{1}+k_{4}^{1}\right)  \frac
{1}{\left(  2m\right)  ^{2}}\int_{\theta}\frac{1}{2\omega}I_{i}(0,\theta
,-\theta,0)\int_{-\infty}^{\infty}dx_{1}^{0}\int_{-\infty}^{0}dx_{2}^{0}%
\int_{-\infty}^{0}dx_{3}^{0}\int_{-\infty}^{0}dx_{4}^{0}\\
&  \times e^{ix_{1}^{0}\left(  k_{1}^{0}-m\right)  +i\left(  x_{2}^{0}%
+x_{1}^{0}\right)  \left(  k_{2}^{0}-\left(  2\omega-m\right)  \right)
+i\left(  x_{3}^{0}+x_{2}^{0}+x_{1}^{0}\right)  \left(  k_{3}^{0}-\left(
m-2\omega\right)  \right)  +i\left(  x_{4}^{0}+x_{3}^{0}+x_{2}^{0}+x_{1}%
^{0}\right)  \left(  k_{4}^{0}+m\right)  }%
\end{align*}
and%
\begin{multline*}
\int_{-\infty}^{\infty}dx_{1}^{0}\int_{-\infty}^{0}dx_{2}^{0}\int_{-\infty
}^{0}dx_{3}^{0}\int_{-\infty}^{0}dx_{4}^{0}e^{ix_{1}^{0}\left(  k_{1}%
^{0}-m\right)  +i\left(  x_{2}^{0}+x_{1}^{0}\right)  \left(  k_{2}^{0}-\left(
2\omega-m\right)  \right)  +i\left(  x_{3}^{0}+x_{2}^{0}+x_{1}^{0}\right)
\left(  k_{3}^{0}-\left(  m-2\omega\right)  \right)  +i\left(  x_{4}^{0}%
+x_{3}^{0}+x_{2}^{0}+x_{1}^{0}\right)  \left(  k_{4}^{0}+m\right)  }\\
=2\pi\delta\left(  k_{1}^{0}+k_{2}^{0}+k_{3}^{0}+k_{4}^{0}\right)  \frac
{-i}{k_{2}^{0}+k_{3}^{0}+k_{4}^{0}+m-i\epsilon}\frac{-i}{k_{3}^{0}+k_{4}%
^{0}+2\omega-i\epsilon}\frac{-i}{k_{4}^{0}+m-i\epsilon}%
\end{multline*}
proves (\ref{G1}) and (\ref{g}). For integrable models typically $S(0)=-1$,
then the contribution from $I_{2}$ vanishes for $\theta_{i}\rightarrow0$ (for
the scaling Ising model we have $S(\theta)\equiv-1$).

\subparagraph{The function $g(x)$ for the scaling Ising model:}

 From (\ref{I}) and (\ref{Fn}) we obtain (up to const)
\begin{align*}
I_{1}(0,\theta,-\theta,0)  &  =\tfrac{1}{4}F(0,\theta-i\pi_{+},-\theta
-i\pi_{-})F(-\theta+i\pi_{+},\theta+i\pi_{-},0)+\left(  \epsilon
\rightarrow-\epsilon\right) \\
&  =\left(  \left(  \tanh\frac{1}{2}\left(  -\theta+i\pi+i\epsilon\right)
\right)  \left(  \tanh\frac{1}{2}\left(  \theta+i\pi-i\epsilon\right)
\right)  \tanh\frac{1}{2}\left(  2\theta\right)  \right) \\
&  \times\left(  \left(  \tanh\frac{1}{2}\left(  -2\theta\right)  \right)
\left(  \tanh\frac{1}{2}\left(  -\theta+i\pi+i\epsilon\right)  \right)
\tanh\frac{1}{2}\left(  \theta+i\pi-i\epsilon\right)  \right)  +\left(
\epsilon\rightarrow-\epsilon\right) \\
&  =\tanh^{2}\theta\coth^{4}\tfrac{1}{2}\left(  \theta-i\epsilon\right)
+\left(  \epsilon\rightarrow-\epsilon\right)  .
\end{align*}

and%
\begin{align}
g(x) &  =-2\pi\int_{\theta}\frac{1}{\omega/m}I_{1}(0,\theta,-\theta,0)\frac
{1}{\omega/m-x}=-\frac{1}{2}\int_{-\infty}^{\infty}d\theta\left(  \frac
{\coth^{4}\tfrac{1}{2}\left(  \theta-i\epsilon\right)  \,\tanh^{2}%
\theta+\left(  \epsilon\rightarrow-\epsilon\right)  }{\cosh\theta\,\left(
\cosh\theta-x\right)  }\right)  \nonumber\\
&  =\frac{16}{1-x}-\frac{15\pi}{2x}-\frac{8}{x}-\frac{4\pi+2}{x^{2}}-\frac
{\pi}{x^{3}}-\frac{\left(  x+1\right)  ^{2}\sqrt{x^{2}-1}}{x^{3}\left(
x-1\right)  ^{2}}2\ln\left(  -x+\sqrt{x^{2}-1}\right)  \label{gk}%
\end{align}
with $g(0)=10\pi+\frac{94}{3}$ (see also \cite{BNNPSW}) and the imaginary part
for $x>1$
\begin{equation}
\operatorname{Im}g(x\pm i\epsilon)=\pm\Theta(x-1)2\pi\frac{\left(  x+1\right)
^{2}\sqrt{x^{2}-1}}{x^{3}\left(  x-1\right)  ^{2}}\,.\label{Img}%
\end{equation}

\subparagraph{The 4-point $\Xi$-function and calculation of $\chi(\omega
_{i},\omega_{f})$:}

\label{here}

The sum over all permutations in (\ref{G1}) yields%
\begin{align*}
{\Xi}(\underline{k})=\frac{1}{32\pi m^{6}}\Bigg\{ &  \left(  \frac{m}%
{k_{4}^{0}+m}+\frac{m}{k_{3}^{0}+m}\right)  \left(  \frac{m}{-k_{1}^{0}%
+m}+\frac{m}{-k_{2}^{0}+m}\right)  g\left(  \frac{-1}{2m}\left(  k_{3}%
^{0}+k_{4}^{0}\right)  \right)  \\
+ &  ~\left(  \frac{m}{k_{4}^{0}+m}+\frac{m}{k_{2}^{0}+m}\right)  \left(
\frac{m}{-k_{1}^{0}+m}+\frac{m}{-k_{3}^{0}+m}\right)  g\left(  \frac{-1}%
{2m}\left(  k_{2}^{0}+k_{4}^{0}\right)  \right)  \\
+ &  ~\left(  \frac{m}{k_{4}^{0}+m}+\frac{m}{k_{1}^{0}+m}\right)  \left(
\frac{m}{-k_{2}^{0}+m}+\frac{m}{-k_{3}^{0}+m}\right)  g\left(  \frac{-1}%
{2m}\left(  k_{1}^{0}+k_{4}^{0}\right)  \right)  \\
+ &  ~\left(  \frac{m}{k_{3}^{0}+m}+\frac{m}{k_{1}^{0}+m}\right)  \left(
\frac{m}{-k_{2}^{0}+m}+\frac{m}{-k_{4}^{0}+m}\right)  g\left(  \frac{-1}%
{2m}\left(  k_{1}^{0}+k_{3}^{0}\right)  \right)  \\
+ &  ~\left(  \frac{m}{k_{2}^{0}+m}+\frac{m}{k_{1}^{0}+m}\right)  \left(
\frac{m}{-k_{3}^{0}+m}+\frac{m}{-k_{4}^{0}+m}\right)  g\left(  \frac{-1}%
{2m}\left(  k_{1}^{0}+k_{2}^{0}\right)  \right)  \\
+ &  ~\left(  \frac{m}{k_{2}^{0}+m}+\frac{m}{k_{3}^{0}+m}\right)  \left(
\frac{m}{-k_{1}^{0}+m}+\frac{m}{-k_{4}^{0}+m}\right)  g\left(  \frac{-1}%
{2m}\left(  k_{3}^{0}+k_{2}^{0}\right)  \right)  \Bigg\}
\end{align*}
Substituting this into (\ref{chi}) we obtain%
\begin{align*}
\chi(\omega_{i},\omega_{f}) &  \sim\frac{\left(  \omega_{i}-\omega
_{f}+2m\right)  ^{2}\operatorname{Im}g\left(  \frac{-1}{2m}\left(  \omega
_{i}-\omega_{f}-i\delta_{12}\right)  \right)  }{\left(  \omega_{i}+m\right)
^{2}\left(  \omega_{f}-m\right)  ^{2}}+\frac{\left(  \omega_{i}+\omega
_{f}+2m\right)  ^{2}\operatorname{Im}g\left(  \frac{-1}{2m}\left(  \omega
_{i}+\omega_{f}-i\delta_{12}\right)  \right)  }{\left(  \omega_{i}+m\right)
^{2}\left(  \omega_{f}+m\right)  ^{2}}\\
&  +\frac{\left(  \omega_{i}+\omega_{f}-2m\right)  ^{2}\operatorname{Im}%
g\left(  \frac{1}{2m}\left(  \omega_{i}+\omega_{f}+i\delta_{12}\right)
\right)  }{\left(  \omega_{f}-m\right)  ^{2}\left(  \omega_{i}-m\right)  ^{2}%
}+\frac{\left(  \omega_{i}-\omega_{f}-2m\right)  ^{2}\operatorname{Im}g\left(
\frac{1}{2m}\left(  \omega_{i}-\omega_{f}+i\delta_{12}\right)  \right)
}{\left(  \omega_{f}+m\right)  ^{2}\left(  \omega_{i}-m\right)  ^{2}}%
\end{align*}
At $\omega_{i}>\omega_{f}>0$ only the last two terms remain and (\ref{r1})
follows with $G(x)=(x-1)^{2}\operatorname{Im}g(x).$

\subparagraph{Proof of (\ref{threeP}):}

We consider the 3 point Greens function
\[
\tau_{\varphi\varphi\epsilon}(\underline{x})=\langle\,0\,|\,T\varphi
(x_{1})\,\varphi(x_{2})\epsilon(x_{3})|\,0\,\rangle
\]
and the Fourier tansform (for $\varphi_{1},\varphi_{2},\varphi_{3}%
=\varphi,\varphi,\epsilon$)%
\begin{align}
\tilde{\tau}_{\varphi\varphi\epsilon}(\underline{k}) &  =\int\underline
{d^{2}x}e^{ix_{i}k_{i}}\tau_{\varphi_{1}\varphi_{2}\varphi_{3}}(\underline
{x})=\sum_{\pi\in S_{3}}\int\underline{d^{2}x}e^{ix_{i}\pi k_{i}}\Theta
_{123}\,\left\langle \varphi_{\pi1}(x_{1})\,\varphi_{\pi2}(x_{2})\varphi
_{\pi3}(x_{3})\right\rangle \nonumber\\
&  =\left(  2\pi\right)  ^{2}\delta^{(2)}\left(  k_{1}+k_{2}+k_{3}\right)
\tilde{\Xi}_{\varphi\varphi\epsilon}(\underline{k}).\label{xi}%
\end{align}
The 3-point Wightman functions in low intermediate particle number
approximation are%
\begin{align*}
w_{\varphi\varphi\epsilon}^{12}(\underline{x}) &  =\frac{1}{2!}\int
_{\underline{\theta}}\langle\,0\,|\,\varphi(x_{1})\,|\,\theta_{1}%
\,\rangle\langle\,\theta_{1}\,|\,\varphi(x_{2})\,|\,\theta_{2},\theta
_{3}\rangle\langle\,\theta_{3},\theta_{2}\,|\,\epsilon(x_{3})\,|\,0\,\rangle\\
w_{\varphi\epsilon\varphi}^{11}(\underline{x}) &  =\int_{\underline{\theta}%
}\langle\,0\,|\,\varphi(x_{1})\,|\,\theta_{1}\,\rangle\,\,\langle\,\theta
_{1}\,|\,\epsilon(x_{2})\,|\,\theta_{2}\rangle\,\langle\,\theta_{2}%
\,|\,\varphi(x_{3})\,|\,0\,\rangle\\
w_{\epsilon\varphi\varphi}^{21}(\underline{x}) &  =\frac{1}{2!}\int
_{\underline{\theta}}\langle\,0\,|\,\epsilon(x_{1})\,|\,\theta_{1},\theta
_{2}\,\rangle\,\langle\,\theta_{1}\,,\theta_{2}|\,\varphi(x_{2})\,|\,\theta
_{3}\rangle\,\langle\theta_{3}\,|\,\varphi(x_{3})\,|\,0\,\rangle.
\end{align*}
As above using the form factor formulas (\ref{e2}), (\ref{Fn}) and the
crossing relations (\ref{cr}) one obtains the Fourier tansforms (for
$k_{i}=(k_{i}^{0},0)$)
\begin{align*}
\tilde{\Xi}_{\varphi\varphi\epsilon}^{12}(k_{1},k_{2},k_{3}) &  =\frac
{-i}{32\pi m^{4}}\frac{m}{k_{1}^{0}-m+i\epsilon}h\left(  -k_{3}^{0}%
/(2m)+i\epsilon\right)  \\
\tilde{\Xi}_{\varphi\epsilon\varphi}^{11}(k_{1},k_{2},k_{3}) &  =-\frac{1}%
{4}\frac{i}{m^{4}}\frac{m}{m-k_{1}^{0}-i\epsilon}\frac{m}{m+k_{3}%
^{0}-i\epsilon}\\
\tilde{\Xi}_{\epsilon\varphi\varphi}^{21}(k_{1},k_{2},k_{3}) &  =\frac
{-i}{32\pi m^{4}}\frac{m}{-k_{3}^{0}-m+i\epsilon}h\left(  k_{1}^{0}%
/(2m)+i\epsilon\right)
\end{align*}
with%
\[
h(x)=\int_{-\infty}^{\infty}\left(  1+\frac{1}{\cosh\theta}\right)  ^{2}%
\frac{1}{\cosh\theta-x}d\theta=-\frac{2}{x}\pi-\frac{2}{x}-\frac{1}{x^{2}}%
\pi-2\frac{x^{2}+2x+1}{x^{2}\sqrt{x^{2}-1}}\ln\left(  -x-\sqrt{x^{2}%
-1}\right)  .
\]
Finally with (\ref{xi}) we obtain
\[
\tilde{\Xi}_{\varphi\varphi\epsilon}(k_{1},k_{2},k_{3})=\tilde{\Xi}%
_{\varphi\varphi\epsilon}^{12}(k_{1},k_{2},k_{3})+\tilde{\Xi}_{\varphi
\epsilon\varphi}^{11}(k_{1},k_{3},k_{2})+\tilde{\Xi}_{\epsilon\varphi\varphi
}^{21}(k_{3},k_{1},k_{2})+\left(  k_{1}\leftrightarrow k_{2}\right)
\]
which proves (\ref{threeP}).


\end{document}